\documentclass[preprint,prb,aps,showpacs]{revtex4}

\usepackage{graphicx}
\def\be{\begin{equation}}
\def\ee{\end{equation}}
\def\bea{\begin{eqnarray}}
\def\eea{\end{eqnarray}}

\begin{document}

\title{Signature of superconducting states in cubic crystal without inversion symmetry}

\author{Chi-Ken~Lu\footnote{
Electronic mail: luck@phys.sinica.edu.tw} and Sungkit~Yip}

\affiliation{Institute of Physics, Academia Sinica, Nankang, Taipei
115, Taiwan}

\date{\today }

\begin{abstract}

{The effects of absence of inversion symmetry on superconducting
states are investigated theoretically. In particular we focus on the
noncentrosymmetric compounds which have the cubic symmetry $O$ like
Li$_2$Pt$_3$B. An appropriate and isotropic spin-orbital interaction
is added in the Hamiltonian and it acts like a magnetic monopole in
the momentum space. The consequent pairing wavefunction has an
additional triplet component in the pseudospin space, and a Zeeman
magnetic field $\bf{B}$ can induce a collinear supercurrent $\bf{J}$
with a coefficient $\kappa(T)$. The effects of anisotropy embedded
in the cubic symmetry and the nodal superconducting gap function on
$\kappa(T)$ are also considered. From the macroscopic perspectives,
the pair of mutually induced $\bf{J}$ and magnetization ${\bf{M}}$
can affect the distribution of magnetic field in such
noncentrosymmetric superconductors, which is studied through solving
the Maxwell equation in the Meissner geometry as well as the case of
a single vortex line. In both cases, magnetic fields perpendicular
to the external ones emerge as a signature of the broken symmetry.}

\end{abstract}


\maketitle

\section{Introduction}

The family of fermion superfluid, which includes the classes of
conventional superconductor, helium-3 superfluid, and cuprate
superconductor, has been one of the most frontier subjects in
condensed matter physics. According to the parity symmetry of
their pairing wavefunction,\cite{symmetry_group} the above classes
can be labeled as $s$-wave, $p$-wave, and $d$-wave superfluid
respectively and each has distinct thermodynamic and transport
properties. In a system without inversion symmetry, this
classification is however invalid, and the system is expected to
simultaneously possess the properties belonging to distinct
classes. Symmetry of the pairing wavefunctions as well as the gap
functions are the immediate question. Theoretical studies based on
the addition of a spin-orbital interaction in the Hamiltonian
predict the Cooper pair to be a mixed state of singlet and triplet
in pseudo-spin space,\cite{Gorkov} which can lead to a
non-vanishing spin susceptibility at zero
temperature.\cite{Gorkov,Edelstein1,spinSus} Besides, the nodal
gap structure has been investigated
experimentally\cite{exp_Ce1,exp_Ce2,exp_Li1,exp_Li2} on two
typical superconducting compounds, CePt$_3$Si and
Li$_2$Pt$_3$B,\cite{Sc_Ce,Sc_Li} which have the point group
symmetries of $C_{4v}$ and $O$, respectively.

On the other hand, the spin-orbital interaction also provides a
correlation between the electric and magnetic degrees of freedom
within the Fermi sea, connecting with the magnetic properties of
the superconducting state in a subtle way. For example, a net
polarization of spins can be induced by a shift of momenta
distribution or vice versa in the superconducting
state.\cite{Edelstein1,Edelstein2,microYip} In other words, the
supercurrent and magnetization can be mutually induced. Therefore,
the macroscopic distributions of current and magnetic field in the
superconducting state can also be used to probe the effect of
lacking inversion
symmetry.\cite{Levitov,MacroYip,Fujimoto,VortexJapan} One should
note that the the form of spin-orbital interaction must vary with
the background crystals of different point group symmetry.
Consequently the magnetic properties pertaining to superconducting
CePt$_3$Si and Li$_2$Pt$_3$B are expected to be quite different.
However, almost all the previous theoretical studies are based on
the symmetry of $C_{4v}$ which allows the Rashba form of
spin-orbital interaction.

In this paper we focus on the magnetic properties pertaining to
the compounds with crystal symmetry of $O$, such as Li$_2$Pt$_3$B.
The starting point is to write down an appropriate spin-orbital
interaction which turns out to act like a magnetic monopole in
momentum space in this case. For simplicity we first consider the
case of isotropic Fermi surface and pairing gap. The supercurrent
${\bf{J}}$ is found to have a component parallel to the applied
Zeeman magnetic field ${\bf{B}}$ and the proportional constant
$\kappa$ is obtained by the linear response theory. For
macroscopic studies, we employ the Maxwell equation to investigate
the distribution of magnetic field in the Meissner geometry and
the case of a single vortex line. Lastly, we also consider the
effects of anisotropy embedded in the cubic symmetry which causes
a power-law dependence of $\kappa(T)$ for very low temperature due
to the appearance of line nodes of superconducting gap functions.

\section{Microscopic derivation of supercurrent induced by a Zeeman field}

The goal of this section is to obtain an expression for the
supercurrent induced by a Zeeman magnetic field in the bulk
superconductor without inversion symmetry. We first consider the
normal state. The lack of inversion symmetry is manifested itself
by the spin-orbital interaction in the Hamiltonian
$H=\sum_{\bf{p}}(H_{\bf{p}})_{\alpha\beta}a^{\dag}_{\bf{p}\alpha}a_{\bf{p}\beta}$,
in which the operator is given by,

\be
    H_{{\bf{p}}}=\frac{p^2}{2m}-E_F-
    \vec{h}_{\bf{p}}\cdot\vec{\sigma}
    \:,\label{H_normal}
\ee and $a_{\bf{p}\alpha}$ are the second-quantized operators for
electron of momentum $\bf{p}$ and spin polarization
$\alpha=\{\uparrow\downarrow\}$ is along the z axis in the
laboratory frame. For convenience we write
$\xi_{\bf{p}}=p^2/2m-E_f$. Note that the spin-orbital interaction
is characterized by the parity-breaking inner product consisting
of a parity-odd $\vec{h}_{\bf{p}}=-\vec{h}_{-\bf{p}}$ and the spin
$\vec{\sigma}$ which is invariant under spatial inversion. It is
convenient to work in the helicity basis (labelled by
$\Uparrow\Downarrow$) in which the operator
$\vec{h}_{\bf{p}}\cdot\vec{\sigma}$ is diagonal, that is,

\be
    \hat{h}_{\bf{p}}\cdot\vec{\sigma}|{\bf{p}}\Uparrow\Downarrow\rangle
    =\pm|{\bf{p}}\Uparrow\Downarrow\rangle\:.
\ee The eigenvalues of $H_{\bf{p}}$ are thus given by
$\epsilon^{\pm}_{\bf{p}}=\xi_{\bf{p}}\mp{h_{\bf{p}}}$ for positive
$\Uparrow$ and negative $\Downarrow$ helicity, respectively. Hence
the degenerate spectrum is split into two branches $\pm$ in the
presence of the spin-orbital interaction. The transformation
between the helicity basis $\Uparrow\Downarrow$ and the laboratory
frame basis $\uparrow\downarrow$ are given by the unitary operator
$U_{\bf{p}}=\exp(-\frac{i}{2}\hat{\bf{k}}\cdot\vec{\sigma}\theta_{\bf{p}})$
which rotate the z axis by an angle of $\theta_{\bf{p}}$ around
the axis of ${\bf{k}}={\bf{z}}\times\hat{h}_{\bf{p}}$. More
explicitly the matrix form of $U_{\bf{p}}$ can be written down in
terms of the coordinate of $\hat{h}_{\bf{p}}$, namely,

\bea
    U_{\bf{p}}=\left(\begin{array}{cccc}
    &\cos\frac{\theta_{\bf{p}}}{2}&-e^{-i\phi_{\bf{p}}}\sin\frac{\theta_{\bf{p}}}{2}\\
    &e^{i\phi_{\bf{p}}}\sin\frac{\theta_{\bf{p}}}{2}&\cos\frac{\theta_{\bf{p}}}{2}
    \end{array}\right)\label{spin_rot}
\eea

Next we include the pairing between two electrons of opposite
momenta on the same branch. In the helicity basis, a general
mean-field description for the pairing potential $H_{\Delta}$ can be
written as

\be
    H_{\Delta}=\sum_{\bf{p}}\left[
    \Delta^*_{+}({\bf{p}})a_{-\bf{p}\Uparrow}a_{\bf{p}\Uparrow}
    +\Delta^*_{-}({\bf{p}})a_{-\bf{p}\Downarrow}a_{\bf{p}\Downarrow}
    +h.c.\right]\:,
\ee where the two gap functions $\Delta_{+}$ and $\Delta_{-}$,
representing the pairing order parameter on the two branches, are
not identical in general. However, the above pairing Hamiltonian
can, by performing the transformation $U$, be restored to the case
of singlet whenever
$\Delta_{\pm}({\bf{p}})=e^{\mp{i}\phi_{\bf{p}}}|\Delta|$. Now the
Nambu representation for the full Hamiltonian $H$ in the helicity
basis can be written as

\bea
    H=\sum_{\bf{p}}\left(\begin{array}{cccc}
    &a^{\dag}_{\bf{p}\Uparrow}&a_{-\bf{p}\Uparrow}
    %a^{\dag}_{\bf{p}\Downarrow}\ a_{-\bf{p}\Uparrow}
    \end{array}\right)
    %\cdot
    \left(\begin{array}{cccc}
    &{\xi_{\bf{p}}-h_{\bf{p}}}&{\Delta_{+}}\\%&0&0\\
    &{\Delta^*_{+}}&{-\xi_{\bf{p}}+h_{\bf{p}}}%&0&0
    \end{array}\right)
    %\cdot
    \left(\begin{array}{cccc}
    &a_{\bf{p}\Uparrow}&\\
    &a^{\dag}_{-\bf{p}\Uparrow}&\\
    %&a_{\bf{p}\Downarrow}&\\
    %&a^{\dag}_{-\bf{p}\Uparrow}&
    \end{array}\right)\nonumber\\+
    \left(\begin{array}{cccc}
    &a^{\dag}_{\bf{p}\Downarrow}&a_{-\bf{p}\Downarrow}
    \end{array}\right)%\cdot
    \left(\begin{array}{cccc}
    &{\xi_{\bf{p}}+h_{\bf{p}}}&{\Delta_{-}}\\%&0&0\\
    &{\Delta^*_{-}}&{-\xi_{\bf{p}}-h_{\bf{p}}}%&0&0
    \end{array}\right)
    %\cdot
    \left(\begin{array}{cccc}
    &a_{\bf{p}\Downarrow}&\\
    &a^{\dag}_{-\bf{p}\Downarrow}&
    \end{array}\right)\:.
\eea

In what follows we employ the method of Matsubara Green's
functions.\cite{review_th1} It is useful to introduce the Nambu
spinor representations $\Psi_{\bf{p}}$ and $\tilde{\Psi}_{\bf{p}}$
for $\Uparrow\Downarrow$ and $\uparrow\downarrow$ basis,
respectively

\be
    \tilde{\Psi}_{\bf{p}}=\left(\begin{array}{cccc}
    &a_{\bf{p}\Uparrow}&\\
    &a_{\bf{p}\Downarrow}&\\
    &a^{\dag}_{-\bf{p}\Uparrow}&\\
    &a^{\dag}_{-\bf{p}\Downarrow}&
    \end{array}\right)\:,
    \Psi_{\bf{p}}=\left(\begin{array}{cccc}
    &a_{\bf{p}\uparrow}&\\
    &a_{\bf{p}\downarrow}&\\
    &a^{\dag}_{-\bf{p}\uparrow}&\\
    &a^{\dag}_{-\bf{p}\downarrow}&
    \end{array}\right)\:.
\ee The Matsubara Green's functions $\check{G}$ in the
$\uparrow\downarrow$ basis are defined in a complex time-ordered
manner as

\be
    G_{\alpha\beta}({\bf{p}},\tau)=-\langle{T}_{\tau}\Psi_{\bf{p}\alpha}(\tau)
    \Psi^{\dag}_{\bf{p}\beta}(0)\rangle\:,
\ee and in the matrix form is

\be
    \check{G}=\left(\begin{array}{cccc}
    &\hat{g}&\hat{f}\\
    &\hat{\bar{f}}&\hat{\bar{g}}
    \end{array}\right)\:,
\ee where $\hat{g}$ and $\hat{f}$ are the matrix forms of the
ordinary Green's functions and Gor'kov Green's functions. We note
that the lower components have the properties
$\hat{\bar{g}}({\bf{p}},\tau)=-\hat{g}^{tr}(-{\bf{p}},-\tau)$ and
$\hat{\bar{f}}({\bf{p}},\tau)=\hat{f}^{\dag}({\bf{p}},-\tau)$. The
Fourier transformation of $G$ is given by

\be
    G_{\alpha\beta}({\bf{p}},\omega_n)=\frac{1}{\beta}
    \int_0^{\beta}d{\tau}e^{i\omega_n\tau}G_{\alpha\beta}({\bf{p}},\tau)\:.
\ee where $1/\beta$ is the temperature and the frequency
$\omega_n$=$(2n+1)\pi/\beta$ is restricted due to the Fermi
statistics. It is easier to first obtain the Green's function
$\check{\tilde{G}}$ by simply inverting the matrix $(i\omega_n-H)$
in the helicity basis. The desired $\check{G}$ can be obtained by
performing the rotation in the pseudospin space using the
followings,

\bea
    \hat{g}({\bf{p}})&=&U_{{\bf{p}}}\hat{\tilde{g}}U^{\dag}_{{\bf{p}}}\:,\\
    \hat{f}({\bf{p}})&=&U_{{\bf{p}}}\hat{\tilde{f}}U_{{-\bf{p}}}^{tr}\:,
\eea where the transformation matrix $U_{{-\bf{p}}}$ for the
opposite momentum is given by
$U_{{\bf{p}}}(-i\sigma_y)e^{i\sigma_z\phi_{\bf{p}}}$. Using the
property that $\sigma_y\vec{\sigma}^{tr}\sigma_y=-\vec{\sigma}$,
the expression for the Green's function $\check{G}$ can be
obtained as follows,

\bea
    \hat{g}&=&\frac{1}{2}\left[(g_{+}+g_{-})
    +(g_{+}-g_{-})\hat{h}_{\bf{p}}\cdot\vec{\sigma}\right]\:,\nonumber\\
    \hat{f}&=&\frac{1}{2}\left[(f_{+}+f_{-})
    +(f_{+}-f_{-})\hat{h}_{\bf{p}}\cdot\vec{\sigma}\right](i\sigma_y)\:.\label{GreensFn}
\eea where the scalar functions $g_{\pm}$ and $f_{\pm}$ are given
below

\bea
    g_{\pm}&=&-\frac{i\omega_n+\epsilon^{\pm}_{\bf{p}}}
    {\omega_n^2+{\epsilon_{\bf{p}}^{\pm}}^2+|\Delta_{\pm}|^2}\:,\nonumber\\
    f_{\pm}&=&\frac{\Delta_{\pm}e^{\pm{i}\phi_{\bf{p}}}}
    {\omega_n^2+{\epsilon_{\bf{p}}^{\pm}}^2+|\Delta_{\pm}|^2}\:.
\eea where we note that the previous condition for the pairing to
recover the singlet is consistent with the condition for which the
triplet component of Gor'kov Green's function vanishes in Eq.\
(\ref{GreensFn}).

In what follows we use the linear response theory to calculate the
supercurrent ${\bf{J}}$ induced by an external Zeeman magnetic
field $\vec{b}$. We express the Fourier-transformed current
operator in terms of the Nambu spinor representation as

\be
    {\vec{\rm{J}}}_{\bf{q}}=\sum_{\bf{p}}
    \Psi^{\dag}_{\bf{p}_{-}\alpha}
    ({\vec{\rm{v}}}_{\bf{p}})_{\alpha\beta}
    \Psi_{\bf{p}_{+}\beta}\:,
\ee where the momentum ${\bf{p}_{\pm}}={\bf{p\pm\frac{q}{2}}}$,
and the velocity operator ${\bf{v}}_{\bf{p}}$ associated with
momentum ${\bf{p}}$ is obtained by taking derivative of
Hamiltonian with respect to ${\bf{p}}$, which gives an identical
result with the previous studies,\cite{Fujimoto}

\be
    \check{\vec{\rm{v}}}_{\bf{p}}%=\nabla_{\bf{p}}H_{\bf{p}}
    =\left(\begin{array}{cccc}
    &\frac{{\bf{p}}}{m}
    -\nabla_{\bf{p}}\vec{h}_{\bf{p}}\cdot\vec{\sigma}&0\\
    &0&\frac{{\bf{p}}}{m}
    +\nabla_{\bf{p}}\vec{h}_{\bf{p}}\cdot\vec{\sigma}^{tr}
    \end{array}\right)
    \:.
\ee The paramagnetic perturbation $V$ resulting from the Zeeman
magnetic field $\vec{b}(r)$ is
$V=-\mu\int{dr}\vec{s}(r)\cdot{\vec{b}}(r)$, here the positive
$\mu$ is the magnetic moment and $\vec{s}(r)$ denotes the local
spin density. $V$ can be represented in Fourier space as
$V=-\mu\vec{s}_{\bf{q}}\cdot{\vec{b}_{-\bf{q}}}$ in which the
Fourier-transformed spin density is given by

\be
    \vec{s}_{\bf{q}}=\sum_{\bf{p}}\Psi^{\dag}_{\bf{p-q}/2}
    \vec{\Sigma}
    \Psi_{\bf{p+q}/2}\:,
\ee where $\vec{\Sigma}$ is the spin operator for the Nambu spinor
representation, given by

\be
    \check{\vec{\Sigma}}=\left(\begin{array}{cccc}
    &\vec{\sigma}&0\\
    &0&-\vec{\sigma}^{tr}
    \end{array}\right)
\ee After some arrangement the current $\vec{\rm{J}}$ can be
written down explicitly as:

\bea
    \vec{{\rm{J}}}({\bf{q}},i\omega_n)
    &=&
    -\int_0^{\beta}d\tau
    e^{i\omega_n\tau}\langle{T_{\tau}}
    \vec{\rm{J}}^{\dag}_{\bf{q}}(\tau)\vec{s}_{\bf{q}}(0)\rangle\cdot(-{\mu\vec{b}_{-\bf{q}}})\nonumber\\
    &=&
    -\frac{\mu}{\beta}
    \sum_{\bf{p}}\sum_{\omega_m}
    \frac{1}{2}\rm{Tr'}\left[\check{\vec{\rm{v}}}_{\bf{p}}\check{G}({\bf{p}}_-,i\omega_m)
    (\check{\vec{\Sigma}}\cdot{\vec{b}_{-\bf{q}}})
    \check{G}({\bf{p}}_+,i\omega_n+i\omega_m)\right]\:,\label{j_first}
\eea in which the symbol $\rm{Tr'}$ denotes taking trace over both
the electron and hole sectors, and a factor of $\frac{1}{2}$ is
added to avoid double counting.

\section{Manifestation of absence of inversion symmetry}

In this section we are going to demonstrate the manifestation of
absence of inversion symmetry in this cubic superconductors from
both microscopic and macroscopic perspectives. Starting from the
general expression for supercurrent in Eq.\ ({\ref{j_first}}), we
investigate the static and homogeneous case, that is, the limits
${\bf{q}}\rightarrow{0}$ and $\omega_n\rightarrow{0}$ are taken.
The resultant static current $\bf{J}$ is collinear with the
applied field, which can be written as

\be
    {\bf{J}}=-\kappa(T)\bf{{B}}\label{j_second}
\ee where the macroscopic magnetic field, or the magnetic
induction is ${\bf{{B}}}=\vec{b}_0$. The appearance of this
coefficient $\kappa$ is an important signature of the lack of
inversion symmetry. In Sec.\ref{micro} $\kappa$ is studied
explicitly for a given isotropic
$\vec{h}_{\bf{p}}$. %The temperature dependence of $\kappa(T)$ at
%very low temperature, $T/\Delta\ll{1}$, is identical to that of
%$1/{\lambda}^2(T)$, where $\lambda$ is the London penetration
%length.

Sec.\ref{Macro} is devoted to the studies of macroscopic aspects,
which deals with the interaction between the magnetic field and
the nonvanishing pairing order parameter $\Delta$. A crucial
addition of Eq.\ (\ref{j_second}) to the ordinary supercurrent and
the corresponding magnetization are the key ingredients for
understanding the new distribution of magnetic field. By the way,
the expressions for $\bf{J}$ and $\bf{M}$ can be obtained by
taking the derivatives of free energy as,\cite{MacroYip}

\bea
    {\bf{J}}&=&2\frac{\delta{F}}{\delta{\vec{\rm{q}}}}\:,\nonumber\\
    {\bf{M}}&=&-\frac{\delta{F}}{\delta{\bf{B}}}\:,
\eea where the gauge-invariant phase gradient
$\vec{\rm{q}}=\hbar\nabla\varphi+\frac{2e}{c}\bf{A}$ and the free
energy $F$ contains an extra term of
$-\frac{1}{2}\kappa\vec{\rm{q}}\cdot\bf{B}$ representing the
absence of inversion symmetry.\cite{MacroYip} More explicitly the
above expressions can be written as

\bea
    \frac{4\pi}{c}e{\bf{J}}&=&\frac{1}{{\lambda}^2}
    ({\bf{A}}+\frac{c\hbar}{2e}\nabla\varphi)
    -\frac{\delta}{{\lambda}^2}{\bf{B}}\:,\label{Max2}\\
    4\pi{\bf{M}}&=&\frac{\delta}{{\lambda}^2}
    ({\bf{A}}+\frac{c\hbar}{2e}\nabla\varphi)\label{Max3}\:.
\eea $\lambda$ denotes the London penetration length. The length
parameter $\delta=\frac{4e\pi}{c}\kappa\lambda^2$ is introduced for
later convenience. Note that we have taken the electronic charge to
be $(-e)$.

\subsection{Microscopic aspects}\label{micro}

Now we consider the simplest case for which the spin orbital
interaction is isotropic, and the gaps are identical for both
branches and isotropic as well, that is,

\bea
    \vec{h}_{\bf{p}}&=&\alpha{\bf{p}}\:,\nonumber\\
    \Delta_{+}&=&\Delta_{-}=\Delta\:.\label{basisFn1}
\eea The strength of spin-orbital interaction is characterized by
the quantity $\alpha$, which has the dimension of velocity and is
weak in the sense that $\alpha/v_F\ll{1}$. Here $v_F$ denotes the
Fermi velocity. Starting from Eq.\ (\ref{j_first}), with some
arrangements, the static current can be obtained by taking the
limit ${{\bf{q}}}\rightarrow{0}$ of the following expression,

\be
    {\bf{J}}_{{\bf{q}}}=-\frac{\mu}{\beta}\int_{-\infty}^{+\infty}
    {d\xi\nu(\xi)}\sum_{\omega_n}\sum_{\mu,\nu=\pm}
    \rm{Q}_{\mu\nu}\frac{(i\omega_n+\epsilon^{\mu}_{\bf{p}_{+}})
    (i\omega_n+\epsilon^{\nu}_{\bf{p}_{-}})+\Delta^2}
    {(\omega_n^2+{\epsilon^{\mu}_{\bf{p}_{+}}}^2+\Delta^2)
    (\omega_n^2+{\epsilon^{\nu}_{\bf{p}_{-}}}^2+\Delta^2)}
    \:,\label{J_third}
\ee where the matrix elements of $\textrm{Q}$ represent the
factors for intra-branch contributions, $\mu=\nu={\pm}$, and the
inter-branch ones, $\mu=-\nu$. Explicitly, $\rm{Q}$ in the matrix
form can be written as

\be
    \rm{Q}=\left(\begin{array}{cccc}
    &\frac{1}{2}\vec{n}-\frac{1}{4}(\vec{l}+\vec{t})&-\frac{1}{4}(\vec{l}-\vec{t})\\
    &-\frac{1}{4}(\vec{l}-\vec{t})&-\frac{1}{2}\vec{n}-\frac{1}{4}(\vec{l}+\vec{t})
    \end{array}\right)\:,
\ee where the three vectors are obtained after the operation of
trace and solid-angle integration, namely,

\bea
    \vec{n}&=&\int\frac{d\Omega}{4\pi}
    \rm{Tr}[\frac{\bf{p}}{m}(\vec{\sigma}\cdot{\bf{B}})
    (\hat{\bf{p}}\cdot\vec{\sigma})]
    =\frac{2}{3}\frac{p}{m}{\bf{B}}\:,\nonumber\\
    \vec{l}&=&\int\frac{d\Omega}{4\pi}
    \rm{Tr}[\alpha\vec{\sigma}(\vec{\sigma}\cdot{\bf{B}})]
    =2\alpha{\bf{B}}\:,\nonumber\\
    \vec{t}&=&\int\frac{d\Omega}{4\pi}
    \rm{Tr}[\alpha\vec{\sigma}(\hat{\bf{p}}\cdot\vec{\sigma})
    (\vec{\sigma}\cdot{\bf{B}})
    (\hat{\bf{p}}\cdot\vec{\sigma})]=-\frac{2}{3}\alpha{\bf{B}}\:.
    \label{nlt}
\eea Note that the trace here is only taken over a two-by-two
helicity space in contrast to the previous operation in Eq.\
(\ref{j_first}). First we note that the static current in Eq.\
(\ref{J_third}) is zero when the spin-orbital interaction is
absent. For $\vec{n}$, $\vec{l}$ and $\vec{t}$ in Eq.\
(\ref{nlt}), therefore, only the contributions up to first order
$\alpha/v_F$ are relevant. Since the summation over $\omega_n$
will give a singular integrant concentrated at the Fermi level, it
is eligible to substitute the quantities $p$ and $\nu(\xi)$ with
their values at Fermi level and then move them out of the
integral. However, the contributions from $\vec{n}$ in the
diagonal parts of $\rm{Q}$ must be taken care of because
explicitly they are of zeroth order of $\alpha/v_F$. Hence the
implicit contributions from modification of Fermi momentum and
density of states due to the spin-orbital interaction have to be
taken into account. Namely, the Fermi momentum for each branch,
$p_F^{\pm}=p_F(1\pm\alpha/v_F)$, and also the density of states at
Fermi level, $\nu^{\pm}=mp_F^{\pm}(1\pm\alpha/v_F)$, should be
used here. We also use the trick which enables performing the
integration of energy first,\cite{QFTbook} and after some algebra
the coefficient $\kappa$ is obtained as

\be
    \frac{\kappa(T)}{\mu\alpha\nu(0)}=-\frac{4}{3}
    \left
    \{
    [
    1-\frac{\pi}{\beta}\sum_n\frac{\Delta^2}
    {(\omega_n^2+\Delta^2)^{\frac{3}{2}}}
    ]%\nonumber\\
    -[1-\frac{\pi}{\beta}\sum_n\frac{1}
    {(\omega_n^2+\Delta^2)^{\frac{1}{2}}}
    \frac{\Delta^2}
    {\omega_n^2+\Delta^2+(\alpha{p_F})^2}]
    \right\}
    \:.\label{kappa}
\ee The term in first bracket is actually the Yoshida function
$Y(\Delta,T)$, which is a universal function characterizing the
single-particle excitation across the gap $\Delta$ at temperature
$T$. For an isotropic gap and at low temperature $T/\Delta\ll{1}$,
the function $Y$ is proportional to $\exp(-\Delta/T)$. The second
bracket is identical to the first one when the spin-orbital
interaction is absent. We denote this term by a function
$y(\Delta,\alpha,T)$ to represent the excitations between two
superconducting states separated by an energy of $\alpha{p_F}$.
Thus we can rewrite Eq.\ (\ref{kappa}) as
$\kappa=\frac{4}{3}\mu\alpha\nu(0)(y-Y)$. At zero temperature, the
function $y$ can be evaluated by replacing the summation over the
Matsubara frequency by an integral, and for isotropic $\Delta$ and
$\alpha$ this function is given by,

\be
    y(\Delta,\alpha,T=0)=
    1-\frac{1}{2\theta}\frac{1}{\sqrt{1+\theta^2}}
    \ln\left(1+2\theta^2+2\theta\sqrt{1+\theta^2}
    \right)\:,\label{y_function}
\ee where the number $\theta=\alpha{p_F}/\Delta$. For small
$\theta$, the function $y\sim{\frac{2}{3}}\theta^2$. For large
$\theta$, it is approximately $1-\frac{\ln(2\theta)}{\theta^2}$.
Both limits coincide with the previous
predictions.\cite{Gorkov,microYip}

In addition, the intra-branch and inter-branch contributions can
be respectively recognized as the Pauli and Van Vleck ones in the
previous studies.\cite{microYip} Therefore the induced current
$\bf{J}$ is absent as the Pauli and Van Vleck contributions cancel
each other in the normal state in which $Y=y=1$. On the other
hand, the existence of such current relies on the fact that the
Pauli paramagnetic contribution in the superconducting state is
significantly suppressed while the Van Vleck one is only reduced
by a small portion as long as $\Delta\ll\alpha{p_F}$. Consequently
one can easily infer that the net supercurrent always flows in
opposite to the Pauli paramagnetic current.

\subsection{Macroscopic aspects}\label{Macro}

Here the effect of lacking inversion symmetry on a macroscopic
length scale is studied through solving the static Maxwell
equation,

\be
    \nabla\times{\bf{B}}=4\pi\nabla\times{\bf{M}}
    +\frac{4\pi}{c}(-e)\bf{J}\:,\label{Max1}
\ee Together with the current and magnetization given by Eq.\
(\ref{Max2}) and (\ref{Max3}), we are able to obtain an equation
in terms of magnetic filed ${\bf{B}}$ only, namely,

\be
    \nabla\times\nabla\times{\bf{B}}=-\frac{1}{\lambda^2}{\bf{B}}
    +2\frac{\delta}{\lambda^2}\nabla\times{\bf{B}}\:,\label{genralBeq}
\ee in which the last curl term is generated from
$\nabla\times\bf{M}$ as well as the collinear supercurrent induced
by $\bf{B}$. Hence, one can expect to observe a transverse component
of the applied Zeeman field in such noncentrosymmetric
superconductors.

Eq.\ (\ref{genralBeq}) can be applied for studying the penetration
of magnetic field in Meissner geometry. Explicitly, we can
consider a cubic superconductor occupying the space for $z>0$. It
is more convenient to first consider a general field
$B_{\rm{x}}(z)\hat{\rm{x}}+B_{\rm{y}}(z)\hat{\rm{y}}$ containing
both x and y components. Consequently, the equation with which the
general field satisfies is,

\be
    \frac{d^2}{dz^2}B_{+}(z)=\frac{1}{\lambda^2}B_{+}(z)
    -2i\frac{\delta}{\lambda^2}\frac{d}{dz}B_{+}(z)\:,
\ee where $B_{+}$ stands for the linear combination
$B_{\rm{x}}+iB_{\rm{y}}$. Defining $B_{+}(z=0^{+})=B_{\rm{in},+}$,
the field just inside the superconductor, the general solution is
then given by,

\be
    B_{+}(z>0)=B_{\rm{in},+}e^{-\frac{z}{\lambda}
    \left(\sqrt{1-\frac{\delta^2}{\lambda^2}}
    +i\frac{\delta}{\lambda}\right)}\:,\label{MeissnerB}
\ee which is identical to the previous results.\cite{Levitov} So
one can expect a slight increase of penetration depth by a factor
of $1/\sqrt{1-\frac{\delta^2}{\lambda^2}}$ for such cubic
superconductors in Meissner geometry. Besides, we note that the
additional oscillation is a consequence of a parallel component of
$\bf{J}$ to $\bf{B}$.

The unknown $B_{\rm{in},+}$ in Eq.\ (\ref{MeissnerB}) can be
determined from the boundary condition
${\bf{B}}_{\rm{ext}}={\bf{B}}_{\rm{in}}-4\pi{\bf{M}}(z=0^{+})$,
which requires the knowledge of magnetization ${\bf{M}}$, or
equivalently the gauge-invariant $\vec{\rm{q}}$. In fact
$\vec{\rm{q}}$ can be obtained from integration of the relation
$(-i)B_{+}(z)=dA_{+}(z)/dz$ with given boundary condition at
infinity. We can assume the homogeneity for phase $\varphi$
throughout the superconductor, which is indicative of vanishing
$\bf{A}$ at $z=\infty$ to ensure zero current there. Consequently
the boundary condition at the surface can be shown to, up to first
order of $\delta/\lambda$, have the following form,

\be
    i\lambda{B}_{\rm{ext},+}=%\frac{1}
    %{\sqrt{1-\frac{\delta^2}{\lambda^2}}}
    A_{\rm{in},+}\:.\label{MeissnerBC}
\ee A similar relation for $B_{-}$ can be obtained from the above
by taking complex conjugates on both sides.

If the external field is ${\bf{B}}_{\rm{ext}}=B\hat{\rm{x}}$, then
$A_{\rm{x}}(0^{+})=0$ as a result of Eq.\ (\ref{MeissnerBC}).
Consequently, $M_{\rm{x}}(0^{+})=0$, which demonstrates that the
parallel field is continuous across the surface. In fact, the
magnetic field $B_{\rm{x}}$ and $B_{\rm{y}}$ inside the
superconductor can be obtained by taking the real and imaginary
parts of Eq.\ (\ref{MeissnerB}), respectively. Up to first order
of $\delta/\lambda$, the two components can be written as,

\bea
    B_{\rm{x}}&=&B\left[\cos\frac{\delta{z}}{\lambda^2}+
            \frac{\delta}{\lambda}
            \sin\frac{\delta{z}}{\lambda^2}
            \right]e^{-\frac{z}{\lambda}}\:,\label{Bx}\\
    B_{\rm{y}}&=&B\left[\frac{\delta}{\lambda}
            \cos\frac{\delta{z}}{\lambda^2}
            -\sin\frac{\delta{z}}{\lambda^2}
            \right]e^{-\frac{z}{\lambda}}\:.\label{By}
\eea Along the direction of applied field, the field $B_x$
penetrates into the superconductor with an additionally slow
oscillation of period about $\lambda/\frac{\delta}{\lambda}$. On
the other hand, $M_{\rm{y}}(0^{+})$ is finite due to the existence
of finite flow velocity proportional to $A_{\rm{y}}$ at the
interface. Hence a discontinuity for field $B_{\rm{y}}$ is
generated at the interface,

\be
    \frac{B_{\rm{in},y}-B_{\rm{ext},y}}{B_{\rm{ext},x}}
    =\frac{\delta}{\lambda}\:,
    %{\sqrt{1-\frac{\delta^2}{\lambda^2}}}\:,
\ee which is different from the previous prediction for
inversion-broken superconductor of $C_{4v}$ symmetry where the
discontinuity happens to the parallel field across the
interface.\cite{MacroYip} The functional form of $B_{\rm{y}}$ in
Eq.\ (\ref{By}) indicates that it has the largest magnitude
$\frac{\delta}{\lambda}B$ at the surface, changes sign at
$z\cong\lambda$ and then decays to zero while slowly oscillating.
Furthermore, the flux associated with the perpendicular
$B_{\rm{y}}$ is zero. This is consistent with the conclusion drawn
from Eq.\ (\ref{MeissnerBC}) that $A_{\rm{x}}=0$ at the interface
since both $A_{\rm{x}}(z=\infty)$ and $A_{\rm{z}}$ are zero.

Eq.\ (\ref{genralBeq}) can also be applied for studying a single
vortex line as a macroscopic signature of lacking inversion
symmetry. We consider the conventional case in which the vortex
line is along the z axis, and the cylindrical coordinates are
adopted here. The components of magnetic field are assumed to be
$B_{\phi}(r)$ and $B_{\rm{z}}(r)$ along the directions of
$\hat{\phi}$ and $\hat{\rm{z}}$, respectively. The $\rm{z}$ and
$\phi$ components of Eq.\ (\ref{genralBeq}) are given by,

\bea
    \left[\frac{1}{r}\frac{d}{dr}(r\frac{d}{dr})
    -\frac{1}{\lambda^2}\right]B_{\rm{z}}(r)&=&
    -\tilde{\kappa}
    \frac{1}{\lambda}
    \frac{1}{r}
    \frac{d}{dr}(rB_{\phi})\:,\label{vortex1}\\
    \left[\frac{d}{dr}(\frac{1}{r}\frac{d}{dr}r)
    -\frac{1}{\lambda^2}\right]B_{\phi}(r)&=&
    \tilde{\kappa}
    \frac{1}{\lambda}
    \frac{d}{dr}B_{\rm{z}}\:,\label{vortex2}
\eea where we denote the dimensionless number
$\tilde{\kappa}=2\delta/\lambda$ for convenience. We can therefore
assume the following perturbation solutions,

\bea
    B_{\rm{z}}&=&B_{\rm{z}}^{(0)}+{\tilde{\kappa}}^2B_{\rm{z}}^{(2)}+...\:,\\
    B_{\phi}&=&\tilde{\kappa}B_{\phi}^{(1)}
    +{\tilde{\kappa}}^3B_{\phi}^{(3)}+...\:.
\eea The zeroth order solution $B_{\rm{z}}^{(0)}$ of Eq.\
(\ref{vortex1}) is just the conventional single vortex line
solution, given by $\frac{\Phi}{2\pi\lambda^2}K_0(r/\lambda)$
where $\Phi$ is a quantum of flux $\frac{\pi\hbar{c}}{e}$ and
$K_0$ is the modified Bessel function of zeroth order. As can be
seen in Eq.\ (\ref{vortex2}), now the transverse field $B_{\phi}$
emerges as a result of the nonzero source proportional to $K_1$
from the identity $K_0'=-K_1$. Up to the first order of
$\tilde{\kappa}$, the transverse field can be written down in
terms of the Green's function $g_1$ associated with Eq.\
(\ref{vortex2}).\cite{BesselRef} Defining the dimensionless
variable $x$ as $r/\lambda$, it can be expressed as,

\bea
    B^{(1)}_{\phi}(x)/{\frac{\Phi}{2\pi\lambda^2}}
    &=&\int_0^{\infty}{x'dx'}g_1(x,x')K_1(x')\:,\nonumber\\
    &=&K_1(x)\int_0^x{x'dx'}I_1(x')K_1(x')
    +I_1(x)\int_x^{\infty}{x'dx'}[K_1(x')]^2\:,
\eea in which $K_1$ and $I_1$ are the modified Bessel functions of
first order. The asymptotic behaviors of the transverse field
distribution are,

\be
    B^{(1)}_{\phi}(x)\sim\left\{
    \begin{array}{cccc}
    &\frac{x}{2}\ln\frac{1}{x}&\:,\ x\rightarrow0\\
    &\sqrt{\frac{\pi{x}}{8}}e^{-x}&\:,\
    x\rightarrow\infty
    \end{array}\right\}\:.
\ee Hence the transverse field increases from zero at the origin,
reaches its maximum at a distance of order $\lambda$ from the
center and is followed by an exponential decay. The above extra
magnetic fields non-collinear with the externally applied one can
in principle be detected by observing the extra precession of
polarized muons when their polarization is parallel to the
external applied field. $\frac{\delta}{\lambda}$ for Li$_2$Pt$_3$B
is of order 10$^{-3}$ using the spin-orbital splitting estimated
by Lee and Pickett.\cite{Lee}

\section{Anisotropic Fermi surface and line nodes of gap}

In previous sections we demonstrate an induced supercurrent
parallel to the external Zeeman field as a signature of lacking
inversion symmetry in cubic superconductors. Actually the
spin-orbital interaction appropriate for the point group $O$
respects all but the elements connected to inversion in $O_h$. The
odd-parity basis functions\cite{YipBasisFn}
$\{p_x^{\rm{n}}\hat{\rm{x}}+p_y^{\rm{n}}\hat{\rm{y}}+p_z^{\rm{n}}\hat{\rm{z}};\rm{n}=1,3,5\}$,
in which the cubic symmetry is embedded, belonging to $A_{1u}$
representation within $O_h$ still can be used to construct the
vectors $\vec{h}_{\bf{p}}$. Similarly, the general gap function
respecting the cubic symmetry,

\be
    \hat{\Delta}({\bf{p}})
    =(\Delta_0({\bf{p}})
    +\vec{\rm{d}}_{\bf{p}}\cdot\vec{\sigma})
    (i\sigma_y)\:,\label{generalGap}
\ee can have the component $\Delta_0({\bf{p}})$ constructed from
the even-parity basis functions belonging to $A_{1g}$
representation while the vector function $\vec{\rm{d}}_{\bf{p}}$,
having identical symmetry properties of $\vec{h}_{\bf{p}}$, can be
constructed from $A_{1u}$ representation. Here both components can
be nonzero since parity is no longer respected. An important
feature of the gap function given in Eq.\ (\ref{generalGap}) is
the possible appearance of zeros when the order parameters
$\Delta_0({\bf{p}})$ and $\vec{\rm{d}}_{\bf{p}}$ can
simultaneously be real after appropriate gauge transformations,
which is true if the time-reversal symmetry is respected in the
system. Consequently it is possible to realize the zeros of gap
function when $|\vec{\rm{d}}_{\bf{p}}|$ exceeds
$|\Delta_0(\bf{p})|$ for some points on the Fermi surface. Gapless
excitation can therefore exist in such superconductors by showing,
for example, a power law temperature dependence of penetration
depth.\cite{Hayashi}

In fact the nodal structure of gap function in the compound
Li$_2$Pt$_3$B was shown to be line nodes through the observation
of linear temperature dependence in penetration depth for very low
temperature.\cite{exp_Li1,Hayashi} Here we shall investigate the
effects of anisotropy and line nodes on the coefficient
$\kappa(T)$ near zero temperature. Since we are only interested in
the regime of very weak Zeeman field, the anisotropy, which could
result in some nonlinear field dependence for stronger field
regime, has little qualitative effect here. Hence only the line
nodes of gap function are relevant to the low temperature behavior
of $\kappa(T)$. The intra-branch, or the Pauli, contributions of
$\kappa(T)$ in Eq.\ (\ref{kappa}) for the isotropic case can be
generalized here by directly replacing the gaps $\Delta_{\pm}$
with $|\Delta_0|\pm|\vec{\rm{d}}|$. As for the inter-branch
contribution, the gaps on both branches are in fact much smaller
than the separation $\alpha{p_F}$ in Li$_2$Pt$_3$B, which suggests
little relevance of actual gap function this contribution.
Moreover, even the appearance of zeros associated with the
spin-orbital interaction $\alpha(\Omega)$ are also irrelevant to
the $T$-dependence of this contribution at very low temperature as
long as the zeros associated with $\alpha$ are not identical to
those associated with the pairing gap.

We thus define a dimensionless quantity
$\gamma(T)\equiv\frac{\kappa(0)-\kappa(T)}{\kappa(0)}$ to present
the temperature dependence due to the line nodes at temperature
close to zero. Furthermore, from previous arguments, only the
intra-branch contributions associated with the pairing gap of
$|\Delta_0|-|\vec{\rm{d}}|$ is significant here. In addition, we
are only interested in the effects due to the line nodes and take
these parameters $\alpha(\Omega)$ and $v_F(\Omega)$ to be
isotropic, which makes the evaluation easier and more accessible.
Next the summation over $\omega_n$ in the function $Y$ in Eq.\
(\ref{kappa}) can be transformed into an integral of energy, which
makes $\gamma(T)$ into the following form,

\be
    \gamma(T)=
    \int\frac{d\Omega}{4\pi}(\cos\theta)^2\Delta_{-}^2(\hat{\Omega})
    \int_0^{\infty}{d\xi}\left(\frac{1}{E^3}\frac{2}{e^{E/T}+1}
    +\frac{1}{2E^2T}\frac{1}{\cosh^2(E/2T)}\right)\:,
\ee where the Zeeman field is assumed to be along the z axis, and
$\Delta_{-}(\hat{\Omega})$ denotes the gap functions of the
direction $\hat{\Omega}$ on the Fermi surface, and
$E=\sqrt{\xi^2+\Delta_{-}^2(\hat{\Omega})}$. We note that the above
integral vanishes for $T$ is exactly zero since $E$ is always
positive for all $\xi$. Hence the contribution for $T$ slightly
larger than zero comes from integration around the solid angles
$\Omega$ associated with the zeros of gap. By the cubic symmetry, we
can infer that there are six sets of line nodes on the Fermi
surface, which as a whole remain invariant under any cubic rotation.
Hence the contributions from the six sets can be divided into
$\gamma(T)=2\gamma_{\parallel}(T)+4\gamma_{\perp}(T)$, where
$\gamma_{\parallel}$ denotes that from the line nodes that are
symmetrically distributed around the z axis, while $\gamma_{\perp}$
denotes remaining sets that are around the x or y axis.

For a given set of line nodes, the gap function can be expanded
around these zeros in the following manner,

\be
    \Delta_{-}(\hat{\Omega})=\Delta'(\theta_c)(\theta-\theta_c)\:,
\ee which means the solid angel $\hat{\Omega}=(\theta,\phi)$
associated with the zeros can be parameterized as
$(\theta_c(\phi),\phi)$ along the azimuthal direction.
$\Delta'(\theta_c)$ denotes the slope of gap function along the
direction of $\hat{\theta}$. Finally the linear temperature
dependence can be extracted from $\gamma(T)$, and the following
expression can be obtained if one extends the upper limit of
$\theta$ integral to infinity.

\bea
    \left(\begin{array}{cccc}&\gamma_{\parallel}(T)&\\
    &\gamma_{\perp}(T)&\end{array}\right)=
    \int\frac{{d\phi}}{2\pi}
    \left(\begin{array}{cccc}&\cos^2\theta_c&\\
    &\sin^2\theta_c\cos^2\phi&\end{array}\right)
    \frac{T}{\Delta'(\theta_c)}\nonumber\\
    \int_{0}^{\infty}\int_{0}^{\infty}x^2dxdy
    \left(\frac{2}{r^3}\frac{1}{e^r+1}+\frac{1}{2r^2}\frac{1}{\cosh^2(r/2)}\right)\:,
\eea in which $x=\theta\Delta'(\theta_c)/T$, $y=\xi/T$ and
$r=\sqrt{x^2+y^2}$. In terms of polar coordinate the integral is
to give $\pi\ln{2}$.

\section{conclusions}

In this work we demonstrate an induced supercurrent parallel to
the external Zeeman magnetic field utilizing the Green's function
method. Besides, the induced supercurrent and the consequent
magnetization modify the distribution of magnetic fields in the
Meissner geometry as well as in the vortex line. Transverse
magnetic fields are generated as a sign of breaking inversion
symmetry in superconductors of point group symmetry $O$ such as
Li$_2$Pt$_3$B.

\begin{center}
{\bf Acknowledgment}
\end{center}

This work is supported by the National Science Council of Taiwan,
R.O.C. under grant No.NSC95-2112-M001-054-MY3.

\end{document}